\documentclass[twocolumn,showpacs,pra]{revtex4}

\usepackage{graphicx}%
\usepackage{dcolumn}
\usepackage{amsmath,amssymb}
\usepackage{bm}

\newcommand{\bfr}{{\bf r}}

\begin{document}

%\preprint{HEP/123-qed}

\title[Short Title]{Surface modes and vortex formation in dilute Bose-Einstein condensates\\ at finite temperatures}% Force line breaks with \\

\author{T.~P. Simula}
% \email{Tapio.Simula@hut.fi}
\author{S.~M.~M. Virtanen}%
\author{M.~M. Salomaa}%
\affiliation{Materials Physics Laboratory, 
Helsinki University of Technology\\
P.~O.~Box 2200 (Technical Physics), FIN-02015 HUT, Finland}

\date{\today}% It is always \today, today, but you may specify any date with \date.

\begin{abstract}
The surface mode spectrum is computed self-consistently for dilute Bose-Einstein condensates, providing the temperature dependence of the surface mode induced vortex nucleation frequency. Both the thermodynamic critical frequency for vortex stability and the nucleation frequency implied by the surface excitations increase as the critical condensation temperature is approached from below. The multipolarity of the destabilizing surface excitation decreases with increasing temperature. The computed finite-temperature critical frequencies support the experimental observations and the zero-temperature calculations for vortex nucleation.

\end{abstract}

%\pacs{PACS here}% PACS, the Physics and Astronomy Classification Scheme.
\maketitle
Since it became feasible to create and observe singly quantized vortices \cite{Matthews1999b,Madison2000a} and vortex lattices \cite{Madison2000b,Abo2001a} in dilute Bose-Einstein condensates, a desire for quantitative understanding how vortices penetrate these unique quantum systems emerged. When the system is externally rotated above a certain critical nucleation frequency $\Omega_\nu$, vortices enter the condensate from its boundary. However, the exact dependence of $\Omega_\nu$ on the external variables and the experimental conditions remain the topic of continuing discussions. The reported minimum ratios between the external rotation and the harmonic trapping frequencies for which vortices have been observed thus far include approximately 0.4 \cite{Haljan2001a} and 0.7 \cite{Hodby2001a} in oblate, and 0.1 \cite{Raman2001a}, 0.3 \cite{Abo2001a}, and 0.7 \cite{Chevy2000a,Madison2001a} in prolate trap geometries. Also, a number of theoretical interpretations have been proposed for explaining these frequencies \cite{Isoshima1999a,Feder1999a,Dalfovo2000a,Feder2001a,Garcia-Ripoll2001a,Mitsubishi,Garcia-Ripoll2001b,Sinha2001a,Anglin2001a,Williams2001a,Garcia-Ripollpreprint,Anglinpreprint,Muryshevpreprint,Kramerpreprint}. 

First, there exists an angular frequency above which the free energy of an axially symmetric vortex state becomes lower than that of the vortex-free state. In most experiments, this thermodynamic critical frequency $\Omega_c$, has been found to be substantially below the lowest frequency for which vortices have been observed. This may be explained in terms of an entry barrier at the surface of the condensate, preventing the transition to the lower free energy state. This, in turn, implies hysteresis effects when the rotating drive is ramped above the nucleation threshold and back again. Hence, vortices should occur in condensates at much lower rotation frequencies than those used to create them. In fact, according to the finite-temperature Popov, G1 and G2 approximations, axisymmetric singly quantized vortex states are predicted to be locally energetically (meta)stable even below $\Omega_c$ \cite{Isoshima1999b,Virtanen2001a} (see, however, the discussion in Ref.~\cite{Virtanen2002a}).

Secondly, there exist nodeless surface excitations, or surfons, in the spectrum of the Bose-Einstein condensate that are localized near the surface of the condensate. These `quantum tidal waves' \cite{Khawaja1999a} are often interpreted as being responsible for the instability of the condensate, leading to the onset of vortex nucleation as they represent a density perturbation on the condensate surface that may evolve into a quantized vortex line, thus changing the topology of the condensate. For rotationally invariant systems, an analog of the Landau criterion \cite{Landau1941} may be expressed for the surfons as
\begin{equation}
\Omega_s=\min_{l}\bigg{\{} \frac{\omega_l}{l} \bigg{\}},
\label{surffreq}
\end{equation}
where $\hbar\omega_l$ is the excitation energy of a surfon carrying the angular momentum $\hbar l$. This angular frequency of the rotating perturbation corresponds to the instability of the vortex-free condensate since beyond it there emerges anomalous negative-energy excitations with high multipolarities in the spectrum of the condensate \cite{Isoshima1999a}. However, the value for the nucleation frequency predicted by Eq.~(\ref{surffreq}) is still too low for serving to explain the observed vortex nucleation thresholds in Refs.~\cite{Chevy2000a,Madison2001a,Hodby2001a}. Accordingly, it has been suggested that in those cases the external perturbation used to rotate the system has mainly driven the ($l=2$) quadrupole surface mode, for which $\omega_l/l$ yields a much higher critical angular frequency. Also, the bending of vortex lines and anomalous vortex core modes have been argued to explain certain aspects of the vortex nucleation process \cite{Garcia-Ripoll2001b,Feder2001a}. 

In this paper, we study finite-temperature effects on vortex nucleation and the thermodynamic stability of vortices in dilute Bose-Einstein condensates. The temperature dependence of the surfon-induced vortex nucleation frequencies, as well as the thermodynamic critical frequency, are computed. The numerical results obtained are consistent with the experimental observations and support the present theoretical understanding of the vortex nucleation process based on zero-temperature field theories. Although our results are in qualitative agreement with those of Ref.~\cite{Mitsubishi}, there exist quantitative differences besides the system parameters, since we do not discard any low-energy quasiparticle excitations from the self-consistent sums, and we do employ the local density approximation taking properly into account also the high-lying quasiparticle states. Furthermore, due to the thermal gas, there is a slight difference between the prediction of Eq.~(\ref{surffreq}) and the frequency at which there emerges anomalous surface mode(s) in the spectrum calculated in the rotating frame of reference.

The stationary generalized Gross-Pitaevskii equation 
\begin{equation}
\mu\phi(\bfr)=\bigg{[}\mathcal{H}_0+ g|\phi(\bfr)|^2 +2g\tilde{n}(\bfr)-\Omega L_z\bigg{]}\phi(\bfr)
\end{equation}
for the macroscopic wavefunction $\phi(\bfr)$ models the Bose-Einstein condensate at finite temperatures. Above, $\mu$ stands for the chemical potential, $\mathcal{H}_0\equiv -(\hbar^2/2M)\nabla^2+V(\bfr)$, and the interaction strength $g=4\pi\hbar^2a/M$, where $a$ is the scattering length for atoms of mass $M$. The wavefunction is normalized according to $\int[|\phi(\bfr)|^2+\tilde{n}(\bfr)]{\rm d}\bfr=N$ and the system is rotated at the angular velocity ${\bm \Omega}$, providing the $N$ trapped atoms with the angular momentum ${\bm L}$. Furthermore, the density of the noncondensate atoms $\tilde{n}(\bfr)$ is determined self-consistently via the relation
\begin{equation} 
\tilde{n}(\bfr)=\sum_q[(|u_q(\bfr)|^2+|v_{q}(\bfr)|^2)n_q +|v_{q}(\bfr)|^2],
\end{equation}
where $n_q=(e^{E_q/k_{\rm B}T}-1)^{-1}$ is the Bose-Einstein distribution function and the quasiparticle amplitudes $u_q(\bfr),v_q(\bfr)$, and eigenenergies $E_q$ are obtained from the Bogoliubov equations 
\begin{subequations}
\begin{eqnarray}
\mathcal{L} u_q(\bfr) +g\phi^2(\bfr)v_q(\bfr)&=&E_qu_q(\bfr), \\ \label{Bogo1}
\mathcal{L}^* v_q(\bfr) +g\phi^{*2}(\bfr)u_q(\bfr)&=&-E_qv_q(\bfr)   \label{Bogo2}.
\end{eqnarray}
\label{Bogo}
\end{subequations}
Within the Popov approximation employed here, $\mathcal{L}\equiv\mathcal{H}_0-\mu +2g[|\phi(\bfr)|^2+\tilde{n}(\bfr)]-\Omega L_z$, and $q$ labels the quasiparticle states whose amplitudes must obey the normalization condition $\int[|u_q(\bfr)|^2-|v_{q'}(\bfr)|^2] {\rm d}\bfr=\delta_{qq'}$. Negative-energy quasiparticle excitations with positive norm are referred to as anomalous modes. The free energy of the system, $F=\langle H_{\rm eff}\rangle -TS$, is evaluated using the equation \cite{Freenote}
\begin{eqnarray}
F=\mu N&-&\frac{g}{2}\int|\phi(\bfr)|^4 {\rm d}\bfr-2g\int\tilde{n}(\bfr)|\phi(\bfr)|^2 {\rm d}\bfr \nonumber \\
&+&\sum_qE_qn_q-\int E_q|v_q(\bfr)|^2 {\rm d}\bfr \\ 
&-&k_{\rm\textsc b}T\sum_q[(1+n_q)\ln(1+n_q)-n_q\ln n_q ].\nonumber 
\label{Free}
\end{eqnarray}
We shall consider cylindrically symmetric states, for which $\phi(\bfr)=\phi(r)e^{im\theta}$, the integer winding number $m$ determining the number of circulation quanta in the vortex. Consequently, the equations may be separated into the radial, angular and axial parts, and the surfons are uniquely labeled by their angular momentum quantum number $l$ (the radial and axial quantum numbers for these states equal zero). The condensate is radially confined by a harmonic trapping potential $V(\bfr)=\frac{1}{2}M\omega_r^2r^2$. In the axial direction we impose periodic boundary conditions. The relevant physics of this system is expected to closest resemble spherical/oblate condensate geometries, as the comparison of the critical frequencies with those in Ref.~\cite{Dalfovo2000a} also suggests. In order to guarantee the accuracy of our results \cite{Numerics}, we compute the discrete quasiparticle states using Eqs.~(\ref{Bogo}) up to $E_q=50\:\hbar\omega_r$ and employ the local density approximation \cite{Giorgini1997a,Reidl1999a} for the remaining states. In the results presented, $\omega_r=2\pi\times 10$ Hz and the effective gas parameter $Na/a_{\sc\rm ho}\approx 90$, where $a_{\sc \rm ho}=\sqrt{\hbar/m\omega_r}$ denotes the harmonic-oscillator length. 
 
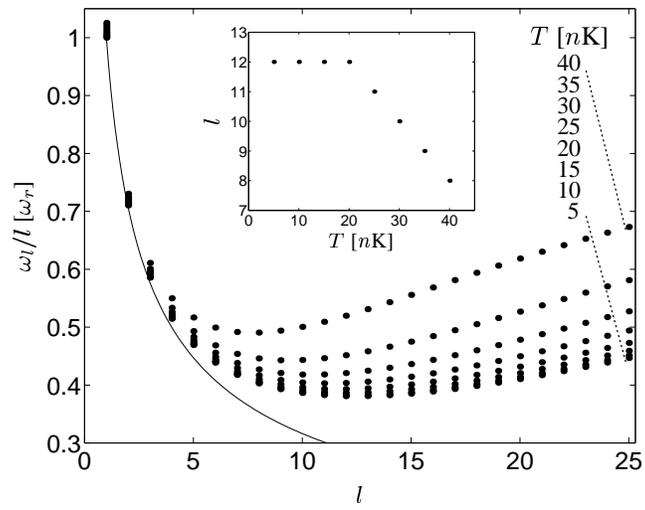
\begin{figure}[h!]
\caption{Temperature dependence of the surface mode energies divided by the corresponding angular momenta as functions of the angular momentum quantum number $l$. The solid line represents the hydrodynamic prediction $1/\sqrt{l}$ \cite{Stringari1996a}. The multipolarity of the destabilizing mode, see Eq.~(\ref{surffreq}), is depicted in the inset as a function of temperature. The critical temperature $T^0_c\approx 45 \;n$K for the parameter values used.}
\label{Fig1}
\end{figure}

Energies per angular momentum of the surface modes with different multipolarities are plotted in Fig.~\ref{Fig1} for a range of temperatures. The solid line is the hydrodynamic prediction $1/\sqrt{l}$ for the surfon dispersion relation \cite{Stringari1996a}. At finite temperatures, Kohn's theorem for parabolic confinement, according to which the dipole mode ($l=1$) should oscillate exactly at the trap frequency $\omega_r$, is not satisfied due to the neglect of the noncondensate dynamics. For all temperatures, the nucleation frequency $\omega_l/l$ implied by the quadrupole mode is found to be close to the value 0.7 observed in the experiments for which the quadrupolar drive may be expected to be the primary source for nucleation \cite{Chevy2000a,Madison2001a,Hodby2001a}. This confirms the negative result found in Ref.~\cite{Hodby2001a} for the significant temperature dependence of the vortex nucleation induced by a quadrupolar drive. That particular nucleation threshold is also expected to be fairly insensitive to the particle number and the geometry of the trap \cite{Dalfovo2000a}. 

However, the energies of the surface modes with higher multipolarities progressively increase at higher temperatures, and the lowest critical frequency $\Omega_s$ due to the surfon excitation is found to increase approximately from 0.4 $\omega_r$ to 0.5 $\omega_r$ with increasing temperature. Also, the corresponding surface mode shifts to a lower value of angular momentum, see the inset in Fig.~\ref{Fig1}, which is a direct consequence of the shrinking of the radius of the condensate.

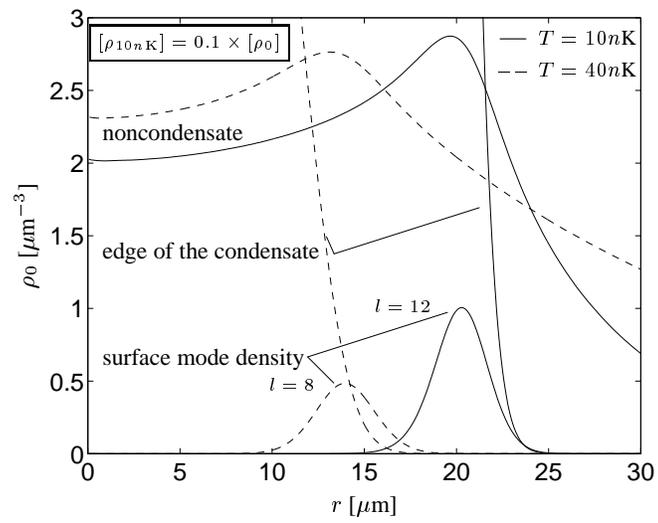
\begin{figure}[h!]
\caption{Densities of the condensate, the noncondensate, and the destabilizing surface mode at $T=40\;n$K($--$) and 10 $n$K (---) as functions of the radial distance from the symmetry axis of the trap. Only part of the condensate boundary is seen in the scope of the figure. For clarity, the densities computed at $T=10\;n$K are scaled up by a factor of 10.}
\label{Fig2}
\end{figure}

The computed radial distributions for the densities of the condensate, noncondensate and the destabilizing surface mode are depicted for $\Omega=0$ in Fig.~\ref{Fig2} for two different temperatures. These surfons are localized near the edge of the condensate, of which only a part is seen within the area of the figure. The shift in the multipolarity of the destabilizing surfon as a function of temperature is illustrated in the figure. As the temperature increases the radial extent of the condensate cloud shrinks. Hence, the angular momentum of the surfon localized in the vicinity of the condensate boundary has the larger angular momentum the lower the temperature is. Notice the noncondensate bulge, due to the mutual repulsion between the condensate and noncondensate particles, on the outer edge of the condensate. 

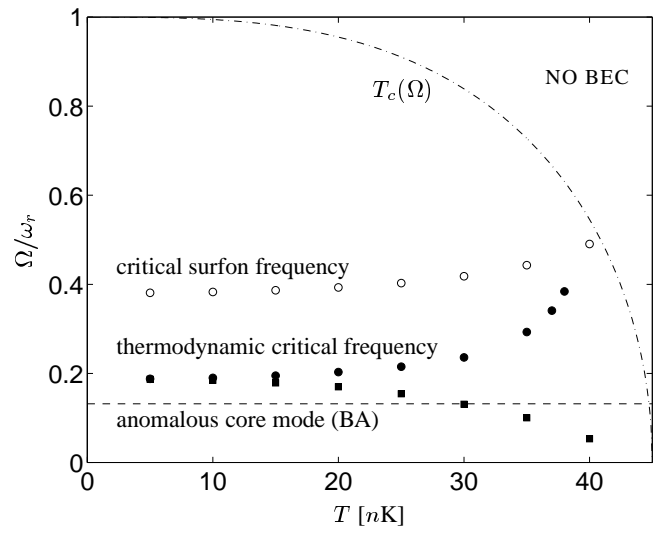
\begin{figure}[h!]
\caption{Thermodynamic ($\bullet$) and minimum surfon-related ($\circ$) critical frequencies for vortex nucleation as functions of temperature. If the rotation of the noncondensate is neglected ({\tiny$\blacksquare$}), the thermodynamic stability of the vortex state increases as the critical temperature is approached from below. Also shown is the critical condensation temperature ($-\cdot-$) and the frequency for local stabilization of the anomalous vortex core mode obtained within the Bogoliubov approximation (BA) ($--$).}
\label{Fig3}
\end{figure}

In Fig.~\ref{Fig3} we have plotted the lowest surfon-related ($\circ$) nucleation frequencies $\Omega_s$ and two different thermodynamic critical frequencies. Also shown is the critical condensation temperature $T_c(\Omega)=T_c^0(1-\Omega^2/\omega^2_r)^{1/3}$ \cite{Stringari1999a} and the nucleation frequency suggested by the anomalous vortex core mode ($--$), calculated within the nonselfconsistent Bogoliubov approximation. Both the thermodynamic and the surfon critical frequencies remain nearly constant below $T_c/2$. At higher temperatures, especially the thermodynamic equilibrium critical frequency $\Omega_c$ $(\bullet)$ strongly deviates from its zero-temperature value, in agreement with the calculation of Ref.~\cite{Stringari1999a}. Furthermore, the critical frequency, given by the energy difference between the vortex and nonvortex states in the absence of rotation ($\Omega=0$), is shown ({\tiny$\blacksquare$}). Physically, this corresponds to the situation in which the noncondensate stays at rest, which has probably been the case in most of the available experiments, excluding those of Ref.~\cite{Haljan2001a}. The decrease of the critical frequency with increasing temperature is probably due to the neglect of the noncondensate dynamics by the Popov approximation whose predictions also for the excitation energies are known to deviate from the experimental values for temperatures $T\gtrsim T_c/2$.

We have also computed the surfon spectrum for an axisymmetric vortex state and found that the corresponding nucleation frequencies are only slightly higher than those presented in Fig.~\ref{Fig3}. This is in accordance with the fact that the angular rotation frequency needed for nucleating vortices grows in proportion to the number of vortices present in the system. Moreover, a presence of multiply quantized vortices in the harmonic trap \cite{Simula2002a} should not affect the qualitative process of vortex formation at the condensate boundary. 

In conclusion, we have computed the surface mode spectrum and the thermodynamic phase diagram for vortex stability in the $T-\Omega$ plane for dilute Bose-Einstein condensates at finite temperatures. According to the current understanding, the highest observed nucleation thresholds with $\Omega/\omega_r \approx 0.7$ can be explained in terms of the excitation of the quadrupole surface modes, whereas the lower values $0.3-0.4$ lie close to the minima of $\omega_l/l$ for the corresponding systems. This scenario for vortex nucleation in dilute Bose-Einstein condensates is suggested to be valid by our computations also at finite temperatures. However, the nucleation value $0.1$ \cite{Raman2001a} coinciding with the thermodynamic critical frequency, cannot solely be explained by the present analysis. A remedy of this observation in the favor of the surfon instability mechanism could be obtained by noting that the local fluid velocity around the small stirrer used in the experiment may well exceed the velocity of the stirrer and thus the effective rotation frequency could be locally much higher \cite{Anglin2001a}.

\begin{acknowledgments}
We thank the Center for Scientific Computing for computer resources,
and the Academy of Finland and the Graduate School in Technical Physics
for support. One of us (TPS) is grateful to the Research Council of Helsinki University of Technology for a postgraduate scholarship. 
\end{acknowledgments}

%\newpage %Just because of unusual number of tables stacked at end
%\bibliography{bibfile}% Produces the bibliography via BibTeX.

\begin{references}
\bibitem{Matthews1999b}
M.~R. Matthews, B.~P. Anderson, P.~C. Haljan, D.~S. Hall, C.~E. Wieman, and E.~A. Cornell, Phys.\ Rev.\ Lett.\ {\bf 83}, 2498 (1999).
\bibitem{Madison2000a}
K.~W. Madison, F. Chevy, W. Wohlleben, and J. Dalibard, Phys.\ Rev.\ Lett.\ {\bf 84}, 806 (2000).
\bibitem{Madison2000b}
K.~W. Madison, F. Chevy, W. Wohlleben, and J. Dalibard, J.\ Mod.\ Opt.\ {\bf 47}, 2715 (2000).
\bibitem{Abo2001a}
J.~R. Abo-Shaeer, C. Raman, J.~M. Vogels, and W. Ketterle,
Science {\bf 292}, 476 (2001).
\bibitem{Haljan2001a}
P.~C. Haljan, I. Coddington, P. Engels, and E.~A. Cornell, Phys.\ Rev.\ Lett.\ {\bf 87}, 210403 (2001).
\bibitem{Hodby2001a}
E. Hodby, G. Hechenblaikner, S.~A. Hopkins, O.~M. Marag\`o, and C.~J. Foot, Phys.\ Rev.\ Lett.\ {\bf 88}, 010405 (2002).
\bibitem{Raman2001a}
C. Raman, J.~R. Abo-Shaeer, J.~M. Vogels, K. Xu, and W. Ketterle,
Phys.\ Rev.\ Lett.\ {\bf 87}, 210402 (2001).
\bibitem{Chevy2000a}
F. Chevy, K.~W. Madison, and J. Dalibard, Phys.\ Rev.\ Lett.\ {\bf 85}, 2223 (2000).
\bibitem{Madison2001a}
K.~W. Madison, F. Chevy, V. Bretin, and J. Dalibard, Phys.\ Rev.\ Lett.\ {\bf 86}, 4443 (2001).
\bibitem{Isoshima1999a}
T.~Isoshima and K.~Machida, Phys.\ Rev.\ A {\bf 60}, 3313 (1999).
\bibitem{Feder1999a}
D.~L. Feder, C.~W. Clark, and B.~I. Schneider, Phys.\ Rev.\ A {\bf 61}, 011601 (1999).
\bibitem{Dalfovo2000a}
F. Dalfovo and S. Stringari,
Phys.\ Rev.\ A {\bf 63}, 011601 (2000).
\bibitem{Feder2001a}
D.~L. Feder, A.~A.~Svidzinsky, A.~L.~Fetter, and C.~W. Clark,
Phys.\ Rev.\ Lett.\ {\bf 86}, 564 (2001).
\bibitem{Garcia-Ripoll2001a}
J.~J. Garc\'{\i}a-Ripoll and V.~M. P{\'e}rez-Garc\'{\i}a,
Phys.\ Rev.\ A {\bf 63}, 041603 (2001).
\bibitem{Mitsubishi}
T. Mizushima, T.~Isoshima and K.~Machida, Phys.\ Rev.\ A {\bf 64}, 043610 (2001).
\bibitem{Garcia-Ripoll2001b}
J.~J. Garc\'{\i}a-Ripoll and V.~M. P{\'e}rez-Garc\'{\i}a,
Phys.\ Rev.\ A {\bf 64}, 053611 (2001).
\bibitem{Sinha2001a}
S. Sinha, and Y. Castin,
Phys.\ Rev.\ Lett.\ {\bf 87}, 190402 (2001).
\bibitem{Anglin2001a}
J.~R. Anglin,
Phys.\ Rev.\ Lett.\ {\bf 87}, 240401 (2001).
\bibitem{Williams2001a}
J.~E. Williams, E. Zaremba, B. Jackson, T.~Nikuni, and A. Griffin,
Phys.\ Rev.\ Lett.\ {\bf 88}, 070401 (2002).
\bibitem{Garcia-Ripollpreprint}
J.~J. Garc\'{\i}a-Ripoll and V.~M. P{\'e}rez-Garc\'{\i}a,
cond-mat/0006368.
\bibitem{Anglinpreprint}
J.~R. Anglin,
cond-mat/0110389.
\bibitem{Muryshevpreprint}
A.~E. Muryshev and P.~O. Fedichev,
cond-mat/0106462.
\bibitem{Kramerpreprint}
M. Kr\"amer, L. Pitaevskii, S. Stringari, and F. Zambelli,
cond-mat/0106524.
\bibitem{Isoshima1999b}
T.~Isoshima and K.~Machida, Phys.\ Rev.\ A {\bf 59}, 2203 (1999).
\bibitem{Virtanen2001a}
S.~M.~M. Virtanen, T.~P. Simula, and M.~M. Salomaa,
Phys.\ Rev.\ Lett.\ {\bf 86}, 2704 (2001).
\bibitem{Virtanen2002a}
S.~M.~M. Virtanen and M.~M. Salomaa,
cond-mat/0203257.
\bibitem{Khawaja1999a}
U. Al Khawaja, C.~J. Pethick, and H. Smith,
Phys.\ Rev.\ A {\bf 60}, 1507 (1999).
\bibitem{Landau1941}
L.~D. Landau,
J.~Phys. (Moscow) {\bf 5}, 71 (1941).
\bibitem{Freenote}
Following I. Kosztin, \v S. Kos, M. Stone, and A.~J. Leggett, Phys.\ Rev.\ B {\bf 58}, 9365 (1998), we define the free energy in terms of the effective mean-field Hamiltonian $H_{\rm eff}$, and the density matrix used in calculating the entropy and the ensemble average is defined by the quasiparticle spectrum of $H_{\rm eff}$.
\bibitem{Numerics}
For details of the numerical methods used in the computations, see Ref.~\cite{Virtanen2001a} and T.~P. Simula, S.~M.~M. Virtanen, and M.~M. Salomaa,
Comput.\ Phys.\ Commun.\ {\bf 142}, 396 (2001).
\bibitem{Giorgini1997a}
S. Giorgini, L.~P. Pitaevskii, and S. Stringari,
J.\ Low Temp.\ Phys.\ {\bf 109}, 309 (1997).
\bibitem{Reidl1999a}
J. Reidl, A. Csord{\'a}s, R. Graham, and P. Sz{\'e}pfalusy,
Phys.\ Rev.\ A {\bf 59}, 3816 (1999).
\bibitem{Stringari1996a}
S. Stringari,
Phys.\ Rev.\ Lett.\ {\bf 77}, 2360 (1996).
\bibitem{Stringari1999a}
S. Stringari,
Phys.\ Rev.\ Lett.\ {\bf 82}, 4371 (1999).
\bibitem{Simula2002a}
T.~P. Simula, S.~M.~M. Virtanen, and M.~M. Salomaa,
Phys.\ Rev.\ A {\bf 65}, 033614 (2002).
\end{references}

\end{document}